\documentclass[twocolumn]{rps-esrl2020}

\usepackage{amsmath,amssymb,amsfonts}
\usepackage{algorithmic}
\usepackage{graphicx}
\usepackage{textcomp}
\usepackage{xcolor}
\usepackage{listings}
\usepackage[linesnumbered,ruled,noend]{algorithm2e}
\usepackage[rflt]{floatflt}
\usepackage[caption=false]{subfig}

\usepackage{tikz}
\usetikzlibrary{automata,backgrounds,calc,fit,positioning,shapes}
\usepackage{pgfplots}
\usepgfplotslibrary{fillbetween}
\usepackage[overlay,absolute]{textpos}
\usepackage{transparent}

\def\papername{\jobname}

\usepackage[square,sort,comma,numbers]{natbib}

\usepackage[color=white]{todonotes}

\usepackage{multirow}

\g@addto@macro\UrlBreaks{\do\-}
\newcommand{\simulink}{\textsc{Simulink}}

\newcommand{\prism}{\textsc{Prism}}
\newcommand{\storm}{\textsc{Storm}}
\newcommand{\errorpro}{\textsc{OpenErrorPro}}

\newcommand{\cG}{\ensuremath{\mathcal{G}}\xspace}

\newcommand{\power}[1]{{\wp}\left(#1\right)}

\newcommand{\Nat}{\mathbb{N}}
\newcommand{\Int}{\mathbb{Z}}

\newcommand{\Guard}{\mathbb{G}}
\newcommand{\AP}{\mathit{AP}}

\newcommand{\Eval}{\mathit{Eval}}
\newcommand{\Var}{\mathit{Var}}

\newcommand{\Prog}{\mathsf{Prog}}
\newcommand{\Bsp}{\mathsf{Bsp}}
\newcommand{\live}{\mathsf{live}}
\newcommand{\wpf}{\mathsf{wp}}
\newcommand{\fail}{\texttt{fail}}
\newcommand{\use}{\mathsf{r}}
\newcommand{\upd}{\mathsf{w}}
\newcommand{\cf}{\mathsf{cf}}
\newcommand{\post}{\mathsf{succ}}
\newcommand{\pred}{\mathsf{pred}}

\newcommand{\lsem}{[\![}
\newcommand{\rsem}{]\!]}

\begin{document}

\markboth{Clemens Dubslaff, Andrey Morozov, Christel Baier, and Klaus Janschek}{
Reduction Methods on Probabilistic Control-flow Programs for Reliability Analysis}
\twocolumn[
\title{Reduction Methods on Probabilistic Control-flow Programs for Reliability Analysis}

\author{Clemens Dubslaff, Andrey Morozov, Christel Baier, and Klaus Janschek}

\address{Technische Universit\"at Dresden, Dresden, Germany.\\
\email{\{clemens.dubslaff, andrey.morozov, christel.baier, klaus.janschek\}@tu-dresden.de}}

\begin{abstract}

Modern safety-critical systems are heterogeneous, complex, and highly dynamic.
They require reliability evaluation methods that go beyond the classical static methods such as fault trees, 
event trees, or reliability block diagrams. Promising dynamic reliability analysis methods employ 
probabilistic model checking on various probabilistic state-based models.
However, such methods have to tackle the well-known state-space explosion problem. 
To compete with this problem, reduction methods such as symmetry reduction and partial-order reduction have been successfully applied to probabilistic models by means of discrete Markov chains or Markov decision processes. Such models are usually specified using probabilistic programs provided in guarded command language. 

In this paper, we propose two automated reduction methods for probabilistic programs that operate on a purely syntactic level: \emph{reset value optimization} and \emph{register allocation optimization}. The presented techniques rely on concepts well known from compiler construction such as live range analysis and register allocation through interference graph coloring. Applied on a redundancy system model for an aircraft velocity control loop modeled in $\simulink$, we show effectiveness of our implementation of the reduction methods. We demonstrate that model-size reductions in three orders of magnitude are possible and show that we can achieve significant speedups for a reliability analysis.
\end{abstract} 

]

\section{Introduction}
Probabilistic model checking (PMC, cf.~\cite{BK08,ForejtKNP11,Katoen16}) 
is an automated technique to analyze stochastic state-based models.
Thanks to extensive tool support, e.g., through probabilistic
model checkers $\prism$~\cite{KNP11} and $\storm$~\cite{DJKV-CAV17},
PMC has been successfully applied to numerous real-world case
studies to analyze systems performance and Quality of Service.
The main issue that hinders PMC to be applicable to large-scale systems
is the so-called \emph{state-space explosion problem} -- the number of states grows
exponentially in the number of components and variables used to model the system.
This problem arises especially when the models are automatically generated
as generators usually do not include techniques of expertise applied during the modeling
process towards handcrafted models~\cite{morozov2019openerrorpro,CDKB18,VolkJK18}.
Thus, reduction techniques have been investigated to reduce the size of the model
while preserving the properties to be analyzed through PMC.
Prominent examples are symmetry reduction~\cite{KNP06a} and probabilistic partial-order
reduction~\cite{BAIER200697}. Also symbolic methods can compete with this problem 
by using concise model representations, e.g., through binary decision diagrams
(BDDs, cf.~\cite{McMillan,BK08}).

Models for PMC tools are usually provided through a probabilistic variant 
of Dijkstra's guarded command language~\cite{Dijkstra1975}, which we call
\emph{probabilistic programs} in this paper. Existing reduction methods 
are exploiting structural properties on the semantics of probabilistic
programs, e.g., on discrete Markov chains (DTMCs)~\cite{Kulkarni} or 
Markov decision processes (MDPs)~\cite{Puterman:book}.

In this paper, we present two reduction methods that operate on a purely 
syntactic level of probabilistic programs:
\emph{reset value optimization (RVO)} and
\emph{register allocation optimization (RAO)}.
To the best of our knowledge, there are no publications yet that report
on automated syntactic reduction methods for probabilistic programs.
For both methods we employ standard concepts from compiler construction
and optimization~\cite{AhoUll77a}
such as \emph{live range analysis} and \emph{register allocation}.
A variable in a program is considered to be \emph{dead}
during a program execution when it not accessed anymore
in a future step. 
The reduction potential then lays in the fact that the value
of dead variables does not influence the future behavior
of the program.
Specifically, RVO rewrites variables to their initial value
as soon as they become dead. In \cite{DDMBJ19} we already
used this concept in a handcrafted fashion to reduce the size of probabilistic 
program models, not exploiting the full reduction potential.
We present a generalized and automated algorithm here in this paper.
RVO is appealing due to its relatively simple concept and implementation.
Differently, RAO is a more sophisticated reduction method that 
relies on register allocation algorithms, i.e.,
algorithms that help to assign a large number of target program variables 
onto a bounded number of CPU registers. Well-known methods are for instance
Chaitin's algorithm \cite{Cha04a} and the linear scan algorithm (LSA)~\cite{poletto1999linear}.
Chaitin's algorithm reduces the problem of register allocation to a
graph-coloring problem (see, e.g., \cite{GJ79}) and is well-known for
its good minimization properties, while the LSA is widely used in compilers
where speed is key (e.g., for run-time compilation).
Within RAO, we use register allocation to determine those variables
in the probabilistic program that can be merged without changing the
program's behavior w.r.t. the properties to be analyzed.
As the main bottle neck of PMC is the state-space explosion problem, we
focus on Chaitin's algorithm in this paper to increase the reduction
potential of RAO.

Our work has been mainly motivated by the need of reduction methods for
automatically generated $\prism$ models for reliability analysis, which
turned out to either being not constructible due to memory constraints
or where a reliability analysis clearly exceeds reasonable time limits~\cite{DDMBJ19}.
While our reduction methods are applicable to a wide range of
probabilistic programs, we present the algorithms in detail for
\emph{probabilistic control-flow programs}, which is an important 
subclass of probabilistic programs that arises in many applications.
Such programs have dedicated control-flow variable storing the current
control-flow location. Although not explicitly stated, many probabilistic
program models analyzed in the PMC community have such a variable, formalizing
discrete time steps or rounds in protocol descriptions.

We formally define RVO and RAO for probabilistic control-flow programs and
present an implementation of both reduction methods, which we evaluate
on a large-scale redundancy system that stems from an aircraft velocity 
control loop (VCL)~\cite{DDMBJ19}. The VCL is based on a
$\simulink$ model~\cite{aircraft,simulink} that is automatically transformed
to a probabilistic control-flow program by the reliability analysis tool
$\errorpro$~\cite{morozov2019openerrorpro}.
Our experiments show that the state space of the VCL model can be reduced
by a factor of 477 by RVO and 1133 by RAO. Together with
symbolic analysis techniques and iterative variable reordering~\cite{DubMorBai20a}
that are required for model construction due to the massive model sizes, 
we also gain speedups for reliability analysis with a factor of 1.6 by RVO and 5.1 by RAO. %

\section{Foundations}\label{sec:preliminaries}
In this section we briefly introduce notations used throughout
the paper, provide notion of programs, and adapt well-known 
live range analysis to this notion.

For a given set $X$ we denote by $\power{X}$ its power set, 
i.e., the set of all subsets of $X$, and by $|X|$ the number
of elements in $X$. Let $X$, $Y$, and $Z$ be
sets with $X\subseteq Y$. For a function $f\colon Y\rightarrow Z$ 
we denote by $f(X)=\{z\in Z: \exists x\in X. f(x)=z\}$.
With $f^{-1}\colon Z\rightarrow\power{Y}$ we denote the inverse function
that given $z\in Z$ returns $\{y\in Y: f(y)=z\}$.
A function $\eta\colon\Var\rightarrow\Int$ over a finite set of variables 
$\Var$ is called a \emph{variable evaluation}.
We denote by $\Eval(\Var)$ the set of variable evaluations over $\Var$.
A \emph{guard} over $\Var$ is a Boolean expression over arithmetic constraints
over $\Var$, e.g., 
\[
	\gamma\enskip =\enskip (x>1)\wedge(y\leq x+5)
\]
is a guard over variables $x,y\in\Var$. By $\Guard(\Var)$ we denote the
set of guards over $\Var$.
Whether a guard is fulfilled in a given variable evaluation is determined
in the standard way, e.g., the guard $\gamma$ above is fulfilled in any
$\eta\in\Eval(\Var)$ with $x,y\in\Var$ where $\eta(x)>1$ and
$\eta(y)\leq \eta(x)+5$. We denote by $\lsem \gamma\rsem\subseteq\Eval(\Var)$
the set of variable evaluations that fulfill $\gamma$.

\subsection{Probabilistic Control-Flow Programs}\label{ssec:programs}
The basis for our reduction methods is provided by 
\emph{probabilistic control-flow programs}, which we
simply also call \emph{programs}. They are a probabilistic version of Dijksta's 
\emph{guarded command language}~\cite{Dijkstra1975}
with a dedicated control-flow mechanism and 
constitute a subset of the input language of the
probabilistic model checker $\prism$~\cite{KNP11}.
Specifically, a \emph{program} is a tuple
\[
	\Prog\enskip = \enskip \big(\, \Var,\, C,\, \iota \,\big)
\]
where $\Var$ and $C$ are finite sets of variables and commands, respectively, 
and $\iota\in\Guard(\Var)$ is a guard that specifies a set of
initial variable evaluations $\lsem\iota\rsem \subseteq \Eval(\Var)$.
We assume a dedicated \emph{control-flow variable} \texttt{cf} 
to be not contained in $\Var$.
A \emph{command} is an expression of the form
\[
\begin{array}{l}
  \texttt{(cf=}\ell\texttt{)}\wedge \texttt{guard} \enskip \rightarrow \\
  \qquad \texttt{p}\textsubscript{1}\texttt{:update}\textsubscript{1} \texttt{ + }
  \ldots  \texttt{ + }
  \texttt{p}\textsubscript{n}\texttt{:update}\textsubscript{n}
\end{array}
\]
where $\ell\in\Nat$ is a control-flow location,
$\texttt{guard}\in\Guard(\Var)$,
and $\texttt{p}\textsubscript{i}\texttt{:update}\textsubscript{i}$
are stochastic updates for $i\in\{1,\ldots,n\}$. 
The latter comprise expressions $\texttt{p}\textsubscript{i}$ that 
describe how likely it is to change variables according to an \emph{update}
\[
	\texttt{update}\textsubscript{i}\enskip = \enskip 
		[\texttt{cf}{:=}\ell',\, x_1{:=}e_1,\, \ldots,\, x_k{:=}e_k].
\]
Here, the update formalizes the change of the control flow
from location $\ell$ to location $\ell'$ and the variables 
$x_j\in\Var$ for $j\in\{1,\ldots,k\}$ are updated with 
the evaluated value of the expression $e_j$. 
In each update, each variable is updated at most once.
We require that for every variable evaluation that fulfills
\texttt{guard} the evaluations of the expressions 
\texttt{p}\textsubscript{1}, \texttt{p}\textsubscript{2}, \ldots,
\texttt{p}\textsubscript{n} constitute a probability distribution,
i.e., sum up to 1. 
The size of a program is the length of a string that arises
from writing all variables, commands, and the initial guard in sequel.

Intuitively, a program defines a state-based semantics where states 
comprise control-flow locations and variable evaluations in $\Eval(\Var)$.
Starting in a control-flow location $\texttt{cf}=0$ and 
variable evaluation $\eta\in\lsem\iota\rsem$,
commands specify how to switch from one state to another:
In case the guard of a command is fulfilled in the current variable
evaluation, the command is considered to be enabled.
Within the set of enabled commands, a fair coin is tossed to
select a command for execution, leading to a transition into a successor state 
by updating variables according to the command's stochastic updates.
To provide an example of stochastic updates, let us consider a
command as above to be enabled in a variable evaluation $\eta\in\Eval(\Var)$.
For some $i\in\{1,...,n\}$, assume $\texttt{p}\textsubscript{i}=x/(1+x)$ and
\begin{center}
	$\texttt{update}\textsubscript{i}\enskip = \enskip [\texttt{cf}{:=}2,\, x{:=}1+y]$
\end{center}	
By executing this command the value of variable $x$ is set to the 
increment of the value of variable $y$ with a probability of $\frac{x}{1+x}$.
With probability $1-\frac{x}{1+x}$, other updates of the command are applied.

In this sense, programs describe stochastic state-based models by
means of discrete Markov chains (DTMCs). 
We denote by $\Pr_{\Prog,\eta}$ the standard probability measure defined over
measurable sets of executions in the DTMC semantics of a program $\Prog$
starting in a variable evaluation $\eta$.
For further details on such models and their use in the context
of verification, we refer to standard textbooks such as \cite{Kulkarni,BK08}. 
Programs used as input of
$\prism$ further specify bounded intervals of integers as domains
of variables, guaranteeing that the arising DTMC is finite.\\

\paragraph{Auxiliary Functions.}
Let $c\in C$ be a command of a program $\Prog$ as defined above. 
We say that a variable is \emph{written}
in $c$ if there is an update for in $c$, and \emph{read} when the
variable appears in some guard, probability expression, or 
used in an expression to update a possibly different variable in $c$.
We denote by $\use,\upd\colon C \rightarrow \power{\Var}$
functions that return those variables $\use(c)$ read in $c$
and those variables $\upd(c)$ written in \emph{every} update of $c$, respectively.

We define a function $\cf\colon C \rightarrow \Int$ that provides
for each command $c$ the control-flow location $\cf(c)$ in which $c$ is enabled.
The functions $\pred,\post\colon C \rightarrow \power{C}$
applied on $c$ return exactly those commands $c'\in\pred(c)$ that have a control-flow 
update to a control-flow location in $\cf(c)$ and those commands 
$c''\in\post(c)$ where $c$ has a control-flow update to some control-flow
location in $\cf(c'')$.

\paragraph{Property Specification.}
Programs as above can be subject to a quantitative analysis, e.g.,
through probabilistic model checking by $\prism$. Usually,
properties are specified in temporal logics over a
set of atomic propositions $\AP$, assuming states 
in the model to be labeled using a function 
$\lambda\colon\AP\rightarrow\Guard(\Var{\cup}\{\texttt{cf}\})$.
For instance, $\fail\in\AP$ could stand for all failure states,
i.e., all $\eta\in\Eval(\Var{\cup}\{\texttt{cf}\})$ that fulfill the label guard $\lambda(\fail)$.
In this paper, we mainly focus on reachability properties such
as $\Diamond\fail$, constituting the set of all program executions
that reach a failure state.
It is well known that such reachability properties are 
measurable in the DTMC semantics of programs (cf. \cite{Kulkarni}).
Bounded reachability properties such as $\Diamond^{< k} \fail$
describe all those executions reaching a failure state within $k$ rounds,
also measurable in this sense~\cite{BDDKK14}.
Here, the number of rounds of an execution of a program is defined
as the number of returning to the initial control-flow location 
$\texttt{cf}=0$.

\begin{example}
\label{bsp:bsp}
Let us consider a program $\Bsp=(\Var,C,\iota)$ where
$\Var=\{x,y\}$, $C$ comprises the commands
\[
\begin{array}{l}
  \texttt{cf}=0\wedge x=1 \enskip\rightarrow\enskip 1:[\texttt{cf}:=1, x:=0]\\
  \texttt{cf}=1\wedge x=0 \enskip\rightarrow\enskip \\
  \enskip 0.5:[\texttt{cf}:=2, y:=0] + 0.5:[\texttt{cf}:=3, y:=0]\\
  \texttt{cf}=2 \enskip\rightarrow\enskip 1:[\texttt{cf}:=0, x:=1]\\
  \texttt{cf}=3 \enskip\rightarrow\enskip \\
  \enskip 0.3:[\texttt{cf}:=0, x:=0] + 0.7:[\texttt{cf}:=1, x:=0]
\end{array}
\]
and $\iota= (x=1 \wedge y=1)$ is the guard
specifying the initial variable evaluation.

Figure~\ref{fig:bsp} shows the DTMC semantics of $\Bsp$,
having seven reachable states. As $\lsem\iota\rsem$ is a
singleton, we only have one initial state $\eta$, highlighted in blue.
Transitions from one state to another are depicted by arrows,
illustrating the step-wise behavior of the program.
There is one deadlock state, i.e., a state without outgoing
transitions, with control-flow location $0$ and a
variable evaluation of $x=0$ and $y=0$.
\begin{figure}
	\vspace*{-1em}	
	\centering
	\includegraphics[scale=.4]{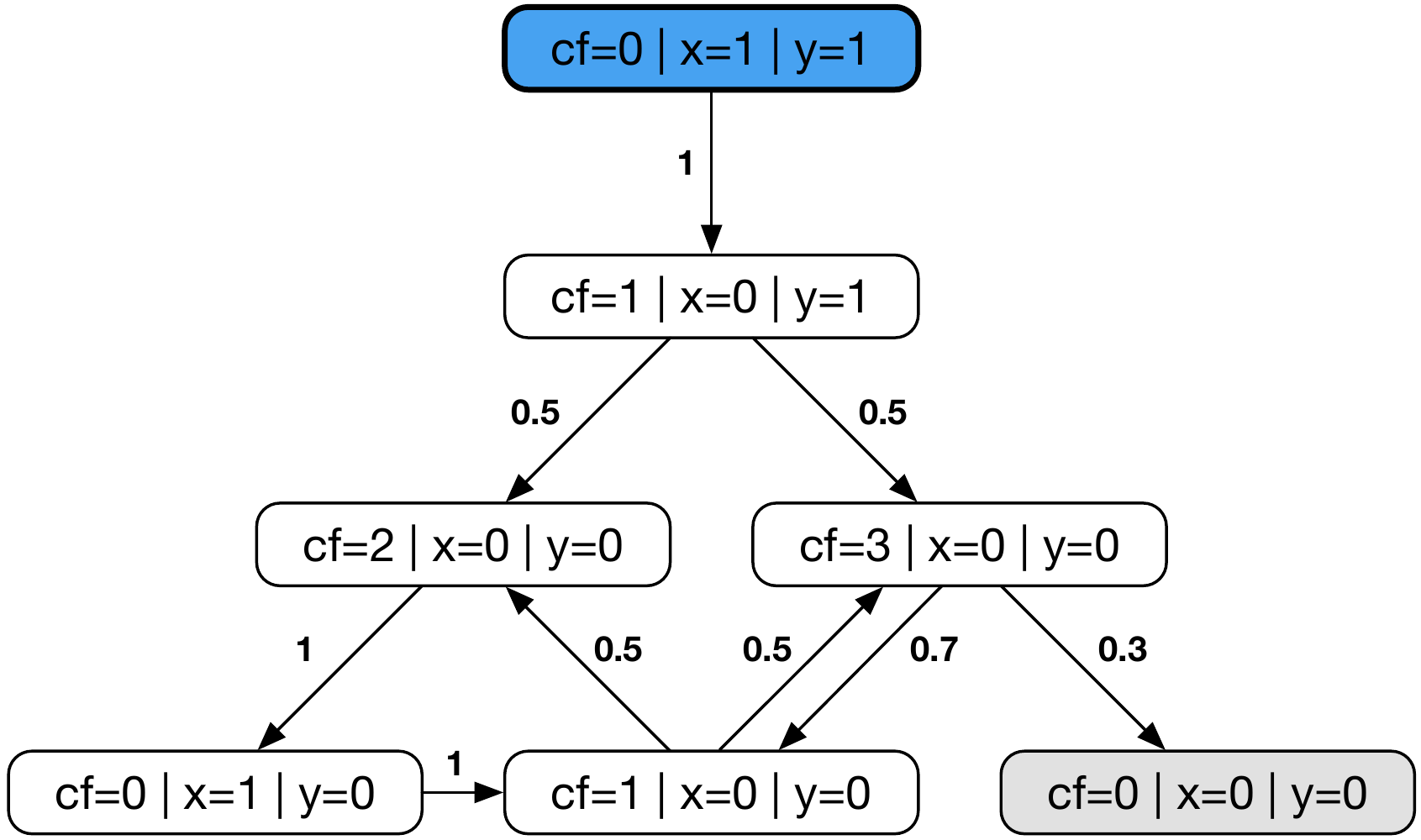}
	\caption{\label{fig:bsp}The DTMC semantics of program $\Bsp$.}
	\vspace*{-1em}	
\end{figure}
A reasonable property to be investigated in  a reliability analysis
would be, e.g., the probability of reaching the deadlock state
within 10 rounds. Formally, this probability would be expressed
by $\Pr\nolimits_{\Prog,\eta}[\Diamond^{< 10}\fail]$ where
$\lambda(\fail)=(\texttt{cf}=0 \wedge x=0)$.
\end{example}

\subsection{Live Range Analysis}
We present here an adaption of the well-known live range analysis
(see, e.g., \cite{AhoUll77a})
to guarded command languages by means of programs.
Intuitively, a variable is considered to be \emph{live} in a 
variable evaluation of the program 
if there is an execution where the variable is not
written until it is read. As we are interested
in reduction methods on a syntactic level, we define an
approach based on commands rather than on variable evaluations.

Assume we have given a program $\Prog$ 
as defined in the last section. 
By Algorithm~\ref{lra-pseudo} we define a function 
$\live\colon C \rightarrow \power{\Var}$
that returns a set of live variables $\live(c)$ for any command $c\in C$.
The function $\live$ is a conservative under-approximation of liveness
based on variable evaluations
in the sense that there could be variable evaluations $\eta$ where a 
command $c$ with $v\in\live(c)$ is enabled but $v$ is not live in $\eta$.
\vspace*{-1em}
\begin{algorithm}[h]
	\SetAlgoLined
	\DontPrintSemicolon
	\SetKwInOut{Input}{input}\SetKwInOut{Output}{output}
	\Input{$\Prog=(\Var,C,\iota)$} 
	\Output{$LRA(\Prog)\colon C \rightarrow\power{\Var}$}
	\BlankLine
	$\live := \use$\;
	$U:=C$\;
	\While{$U\neq \varnothing$}{
		Pick a command $c' \in U$\;
		$U:= U\setminus\{c'\}$\;
		\ForAll{$c\in \pred(c')$}{
			$I:=\live(c)\cup\big(\live(c')\setminus\upd(c)\big)$\;
			\If{$I \neq \live(c)$}{$\live(c):= I$\; $U:=U\cup\{c\}$}
		}
	}     
	\Return $\live$ \;	
	\caption{Live range analysis ($LRA$)}
	\label{lra-pseudo}
\end{algorithm}
\vspace*{-1em}
It is easy to see that Algorithm~\ref{lra-pseudo} terminates
as in every iteration of the while loop (line 3-10) 
either at least one new live variable is added to a command
or the size of $U$ decreases. Furthermore,
Algorithm~\ref{lra-pseudo} runs in polynomial time in the size of the input program.
\vspace{-.5em}
\subsection{Graph Coloring}
Let us revisit notions for coloring graphs.
A \emph{graph} is a pair $\cG=(V,E)$ where $V$ is a finite set of
vertices $V$ and $E\subseteq \power{V}$ is a set of edges where
for each $e\in E$ we have $|e|=2$. Given a graph $\cG=(V,E)$ and 
an integer $k\in\Nat$, a function $f\colon V\rightarrow \{1,\ldots,k\}$ 
is a \emph{$k$-coloring} of $\cG$ if for all $\{x,y\}\in E$ we have 
that $f(x)\neq f(y)$. We call $k\in\Nat$ for which there is a 
$k$-coloring of $\cG$ \emph{minimal} if there is no $k'\in\Nat$ 
with $k'<k$ where there is a $k'$-coloring of $\cG$.
It is well known that the question whether some $k\in\Nat$ is
minimal is NP-hard~\cite{GJ79}. 
To this end, many heuristic have been proposed to determine
nearly minimal graph colorings in reasonable time.
In this paper we rely on a greedy heuristics for graph
coloring that runs in quadratic time and is due to 
Welsh and Powell~\cite{Welsh1967AnUB},
listed in Algorithm~\ref{algo:welshpowell}.
\vspace*{-1.5em}
\begin{algorithm}[h]
	\SetAlgoLined
	\DontPrintSemicolon
	\SetKwInOut{Input}{input}\SetKwInOut{Output}{output}
	\Input{$\cG=(V,E), V=\{1,\ldots,n\}$} 
	\Output{$\mathit{WP}(\cG)\colon V\rightarrow\{1,\ldots,n\}$}
	\BlankLine
	\ForAll{$v\in V$}{
		$f(v):=0$\;
		$d(v):=|\{\{v,v'\}\in E: v'\in V\}|$
	}
	Let $o\colon \{1,\ldots,n\}\rightarrow V$ such that 
		$d(o(i))\geq d(o(j))$ for all $1{\leq}i{\leq}j{\leq}n$\;
	\For{$c := 1,\ldots,n$}{
	\For{$i := 1,\ldots,n$, $f(o(i))=0$}{
		\If{$\nexists v.\{o(i),v\}\in E\wedge f(v){=}c$}
		{$f(o(i))=c$}
	}}
	\Return $f$ \;	
	\caption{Welsh-Powell algorithm}
	\label{algo:welshpowell}
\end{algorithm}
\vspace*{-1.5em} %
\vspace*{-1em}
\section{Reset Value Optimization}\label{sec:reset}
The basic idea behind \emph{reset value optimization 
(RVO)}~\cite{GarSer06a,DDMBJ19}
is to update variables with a default value as soon as they are
not live anymore. 
This approach has the advantage of being relatively
easy to apply on programs even when performed by hand, as after
a simple live range analysis it mainly operates locally on commands.
Even though it is surprisingly effective.

Algorithm~\ref{rvo-pseudo} shows the pseudo code of RVO. %
\vspace*{-1em}%
\begin{algorithm}[h]
	\SetAlgoLined
	\DontPrintSemicolon
	\SetKwInOut{Input}{input}\SetKwInOut{Output}{output}
	\Input{$\Prog=(\Var,C,\iota)$; 
		$\rho\in\Eval(\Var)$; $Ex\subseteq\Var$} 
	\Output{$\Prog'=(\Var,C',\iota)$}
	\BlankLine
	$\live := \mathit{LRA}(\Prog)$\;
	$C':=\varnothing$\;
	\ForAll{$c \in C$}{
		$c' = c$\;
		\ForAll{\texttt{p:}$[$\texttt{cf}:=$\ell, xu]$ in $c$}{
			$L := \live(c){\setminus}Ex$\;
			\ForAll{$s \in C$ with $\cf(s)=\ell$}{
				$L := L{\setminus}\live(s)$
			}
			\ForAll{$x\in L$}{
				\eIf{there is $x:=e$ in $xu$}{replace $x:=e$ in $xu$ of $c'$ by $x := \rho(x)$}
				{add $x := \rho(x)$ to $xu$ in $c'$}
			}
		}
		$C' := C'\cup\{c'\}$\;
	}
	\Return $(\Var,C',\iota)$ \;	
	\caption{RVO}
	\label{rvo-pseudo}
\end{algorithm}
\vspace*{-1em}%
Besides a program $\Prog$
as introduced in Section~\ref{ssec:programs}, it takes a 
variable evaluation $\rho$ and a set of variables $Ex$ as input.
While $\rho$ specifies the values the variables are reset to,
$Ex$ specifies the variables that should be excluded from the
optimization, i.e., not reset. 
Exclusion of variables is required to guarantee 
bisimilarity~\cite{LarSko91a} w.r.t. atomic-proposition
labeling used in properties for analyzing the system.
For instance, assume a variable $v$ that is initially $0$ and 
is set to $1$ when a failure $\fail$ occurred. In case $v$ is set to $1$
in the first round of executing $\Prog$, never 
read again, and no further failure occurs, this variable would
eventually be reset to $0$ in a program $\Prog'$ that results from RVO
on $\Prog$. However, it might be that $\Pr_{\Prog,\eta} [ \Diamond^{< 2} \fail]\neq 
\Pr_{\Prog',\eta} [ \Diamond^{< 2} \fail]$ as an execution fulfilling 
$\Diamond^{< 2}\fail$ in $\Prog$ might be not present in $\Prog'$.

Algorithm~\ref{rvo-pseudo} iterates through all updates in commands of the program
and modifies them depending on a live range analysis (line 5-13):
First, the variables that are live in the command's updates
and can serve as reset candidates are stored in a set $L$ (line 6). Then, all live
variables in commands that have a control-flow location that is set 
by the update are disregarded from $L$ (line 7-8), as these are those
variables that are further live after the update and hence, the value
of these variables matter in future steps. When there are any variables
$x$ in $L$ after the removal of future live variables, then these variables
can be set to any value without changing the behavior of the program.
This is done in line 9 and following. In case there is already an update
of the variable $x$ in the list of updates $xu$, then this update
is replaced by $x:=\rho(x)$ (line 11), and otherwise $x:=\rho(x)$ is added to
the list of updates $xu$ (line 13).
It is easy to see that Algorithm~\ref{rvo-pseudo} terminates and
runs in polynomial time as $C$ is 
finite and $C'$ strictly increases every step until $|C'|=|C|$.
\begin{theorem}
Let $\Prog=(\Var,C,\iota)$ a program, $Ex\subseteq\Var$, $\fail\in\Guard(Ex)$, 
$k\in\Nat$, and $\Prog'$ the program obtained from applying Algorithm~\ref{rvo-pseudo} 
on $\Prog$, $Ex$, and some $\rho\in\Eval(\Var)$. Then for every $\eta\in\lsem\iota\rsem$
\begin{center}
	$\Pr\nolimits_{\Prog,\eta}[\Diamond^{< k}\fail] \enskip=\enskip
	\Pr\nolimits_{\Prog',\eta}[\Diamond^{< k}\fail]$.
\end{center}
\end{theorem}

\begin{example}
\label{bsp:rvo}
Let us return to our example program introduced in Example~\ref{bsp:bsp}.
After applying Algorithm~\ref{rvo-pseudo} on $\Bsp=(\Var,C,\iota)$ with 
$Ex=\varnothing$ and $\rho\in\lsem\iota\rsem$
(recall that $\rho$ is uniquely defined as $\lsem\iota\rsem$ is a singleton)
we obtain a program $\Bsp'=(\Var,C',\iota)$ where $C'$ comprises commands
\[
\begin{array}{l}
  \texttt{cf}=0\wedge x=1 \enskip\rightarrow\enskip 1:[\texttt{cf}:=1, x:=0]\\
  \texttt{cf}=1\wedge x=0 \enskip\rightarrow\enskip \\
  \enskip 0.5:[\texttt{cf}:=2, \textcolor{red}{x:=1}, y:=0]\ + \\
  \enskip 0.5:[\texttt{cf}:=3, \textcolor{red}{x:=1}, y:=0]\\
  \texttt{cf}=2 \enskip\rightarrow\enskip 1:[\texttt{cf}:=0, x:=1, \textcolor{red}{y:=1}]\\
  \texttt{cf}=3 \enskip\rightarrow\enskip \\
  \enskip 0.3:[\texttt{cf}:=0, x:=0] + 0.7:[\texttt{cf}:=1, x:=0]
\end{array}
\]
In the above commands, we highlighted the changes of the updates in red.
According to the program logic, the variable $x$ is not used after the execution of 
the command that is enabled in control-flow location $\texttt{cf}=1$,
since $x$ is not in any guard of commands with control-flow locations 
$\texttt{cf}=2$ and $\texttt{cf}=3$ where in both cases the value of $x$
is updated. To this end, we can explicitly reset it to the 
initial value $1$ without changing the program behavior. Likewise,
variable $y$ can be reset to $1$ in the command with $\texttt{cf}=2$, as $y$
is not part of the guard of the command with $\texttt{cf}=0$ and is updated
in the command of $\texttt{cf}=1$.
Figure \ref{fig:bres} shows the DTMC semantics of $\Bsp'$.
\begin{figure}
	\vspace*{-1em}	
	\centering
	\includegraphics[scale=.4]{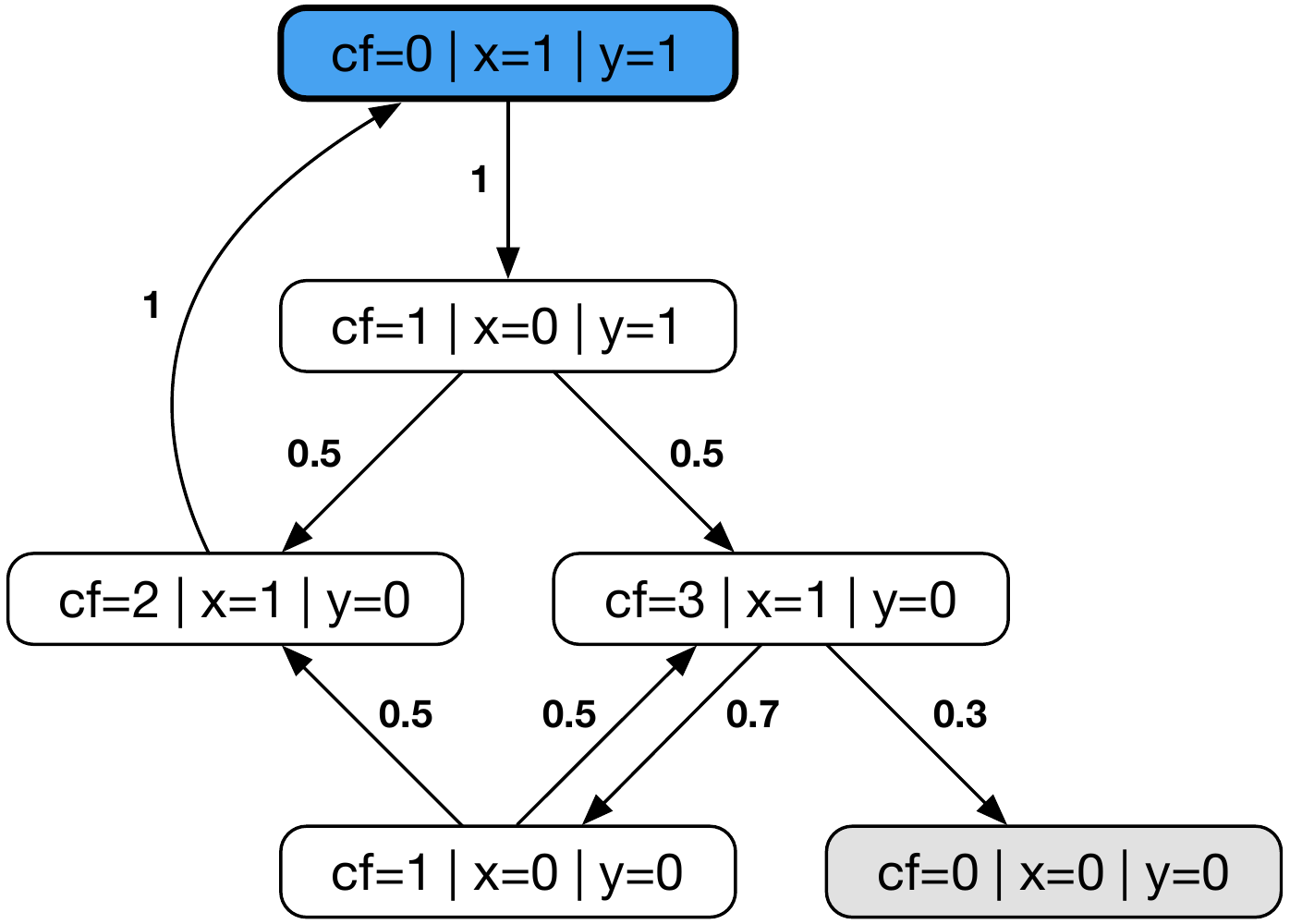}
	\caption{\label{fig:bres}The DTMC semantics of program $\Bsp$ after RVO.}
	\vspace*{-1em}
\end{figure}
While the DTMC semantics of $\Bsp$ had seven states, the DTMC
of Figure \ref{fig:bres} has only six states, witnessing the
potential of reduction using RVO.
\end{example}

\section{Register Allocation Optimization}\label{sec:register}
Ideas from register allocation~\cite{AhoUll77a} can be used
to reduce the number of variables in a program specification
without changing the program's properties, i.e., strive towards
a bisimilar program of reduced size.
Within register allocation, the task is to assign variables to
a bounded number of resources in terms of memory cells.
In contrast to classical register-allocation algorithms,
we do not consider a strict bound on the number of memory cells 
but want to minimize its number, hence not relying on spilling 
techniques. We adopt Chaitin's algorithm~\cite{Cha04a}
known for its good minimizations based on coloring interference graphs.

\subsection{Interference Graph Coloring}
Let $\Prog=(\Var,C,\iota)$ a program. The \emph{interference graph (IG)}
	of $\Prog$ is a graph $\cG_\Prog=(\Var, E)$ where
	$e=\{x,y\}\in E$ iff 
	\begin{itemize}
		\item[(a)] $e \subseteq \live(\cf^{-1}(0))$ or 
		\item[(b)] there is a $c\in C$ with $e\subseteq\live\big(\post(c)\big)$.
	\end{itemize}
Stated in words, two variables in the IG are connected
in case they are not simultaneously live in either the initial
commands or while executing a command.
This intuitively means that one cannot use the same memory cell to store
both variables, i.e., cannot \emph{merge} them.
The construction of IG is doable in time polynomial in
the size of $\Prog$.

To obtain sets of variables that can be merged, the idea is to determine a
coloring of the IG and merge $x,y\in\Var$ in case $x$ and $y$ have
the same color. For obtaining a graph coloring with a nearly 
minimal number of colors, we rely on the Welsh-Powell heuristics 
provided in Algorithm~\ref{algo:welshpowell}.

\subsection{Variable Merging}
Algorithm~\ref{rao-pseudo} shows our register allocation optimization
based on a given coloring of the program's IG.
As within RVO presented in Algorithm~\ref{rvo-pseudo},
this algorithm takes as input also a set of variables $Ex\subseteq\Var$ to be excluded
from optimization to ensure bisimilarity w.r.t. atomic 
propositions used in property specifications.
\vspace*{-1.5em}
\begin{algorithm}[h]
	\SetAlgoLined
	\DontPrintSemicolon
	\SetKwInOut{Input}{input}\SetKwInOut{Output}{output}
	\Input{$\Prog=(\Var,C,\iota)$; $Ex\subseteq\Var$} 
	\Output{$\Prog'=(\Var',C',\iota')$}
	\BlankLine
	$\wpf:=\mathit{WP}(\cG_\Prog)$\;
	$\Var'= Ex \cup \{v_i: \exists v,i. \wpf(v)=i\}$\;
	$C':=C$, $\iota':=\iota$\;
	\ForAll{$v\in\Var{\setminus}Ex$}{
		replace $v$ in $C'$ and $\iota'$ by $v_{\wpf(v)}$\;
	}
	\Return $(\Var',C',\iota')$ \;	
	\caption{RAO}
	\label{rao-pseudo}
\end{algorithm}
\vspace*{-1.5em}

In line 1, we apply Algorithm~\ref{algo:welshpowell} on the IG of 
the input program $\Prog$. Line 2 specifies the new set of variables
$\Var'$ as the excluded variables $Ex$ from $\Var$ (to guarantee correctness
of analysis results w.r.t. properties over $\Guard(Ex)$)
and merged variables. The rest of the algorithm simply replaces any
occurrence of variables not excluded from optimization by
their merged counterparts. Algorithm~\ref{rao-pseudo} runs in polynomial
time as its complexity is dominated by the live range analysis 
(see Algorithm~\ref{lra-pseudo}), the construction of IG, 
and the Welsh-Powell algorithm (see Algorithm~\ref{algo:welshpowell}).

\begin{theorem}
Let $\Prog=(\Var,C,\iota)$ a program, $Ex\subseteq\Var$, $\fail\in\Guard(Ex)$, $k\in\Nat$, 
and $\Prog'$ the program obtained from applying Algorithm~\ref{rao-pseudo} 
on $\Prog$ and $Ex$. Then for every $\eta\in\lsem\iota\rsem$
\begin{center}
	$\Pr\nolimits_{\Prog,\eta}[\Diamond^{< k}\fail] \enskip=\enskip
	\Pr\nolimits_{\Prog',\eta}[\Diamond^{< k}\fail]$.
\end{center}
\end{theorem}
\vspace{-.5em}
\begin{example}
\label{bsp:rao}
Let us again return to our example program introduced in Example~\ref{bsp:bsp}.
After applying Algorithm~\ref{rao-pseudo} on $\Bsp=(\Var,C,\iota)$ with 
$Ex=\varnothing$ we obtain a program $\Bsp'=(\Var',C',\iota')$ where
$\Var=\{xy\}$, $\iota'=(xy=1)$, and $C'$ comprises the commands
\[
\begin{array}{l}
  \texttt{cf}=0\wedge \textcolor{red}{xy=1} \enskip\rightarrow\enskip 1:[\texttt{cf}:=1, \textcolor{red}{xy:=0}]\\
  \texttt{cf}=1\wedge \textcolor{red}{xy=0} \enskip\rightarrow\enskip \\
  \enskip 0.5:[\texttt{cf}:=2, \textcolor{red}{xy:=0}]\ + \\
  \enskip 0.5:[\texttt{cf}:=3, \textcolor{red}{xy:=0}]\\
  \texttt{cf}=2 \enskip\rightarrow\enskip 1:[\texttt{cf}:=0, \textcolor{red}{xy:=1}]\\
  \texttt{cf}=3 \enskip\rightarrow 
  \enskip 0.3:[\texttt{cf}:=0, \textcolor{red}{xy:=0}]\ + \\
  \hspace{5.58em} 0.7:[\texttt{cf}:=1, \textcolor{red}{xy:=0}]
\end{array}
\]
Again, we highlighted the changes of the commands compared to $\Bsp$ in red.
Already in Example~\ref{bsp:rvo}, we saw that $x$ and $y$ do not interfere
in resets and the use of them in guards. It is hence not too surprising that
we can merge them into one variable $v_1=xy$ according to Algorithm~\ref{rao-pseudo}.
Figure \ref{fig:breg} shows the DTMC semantics of $\Bsp'$.
\begin{figure}
	\vspace*{-1em}	
	\centering
	\includegraphics[scale=.4]{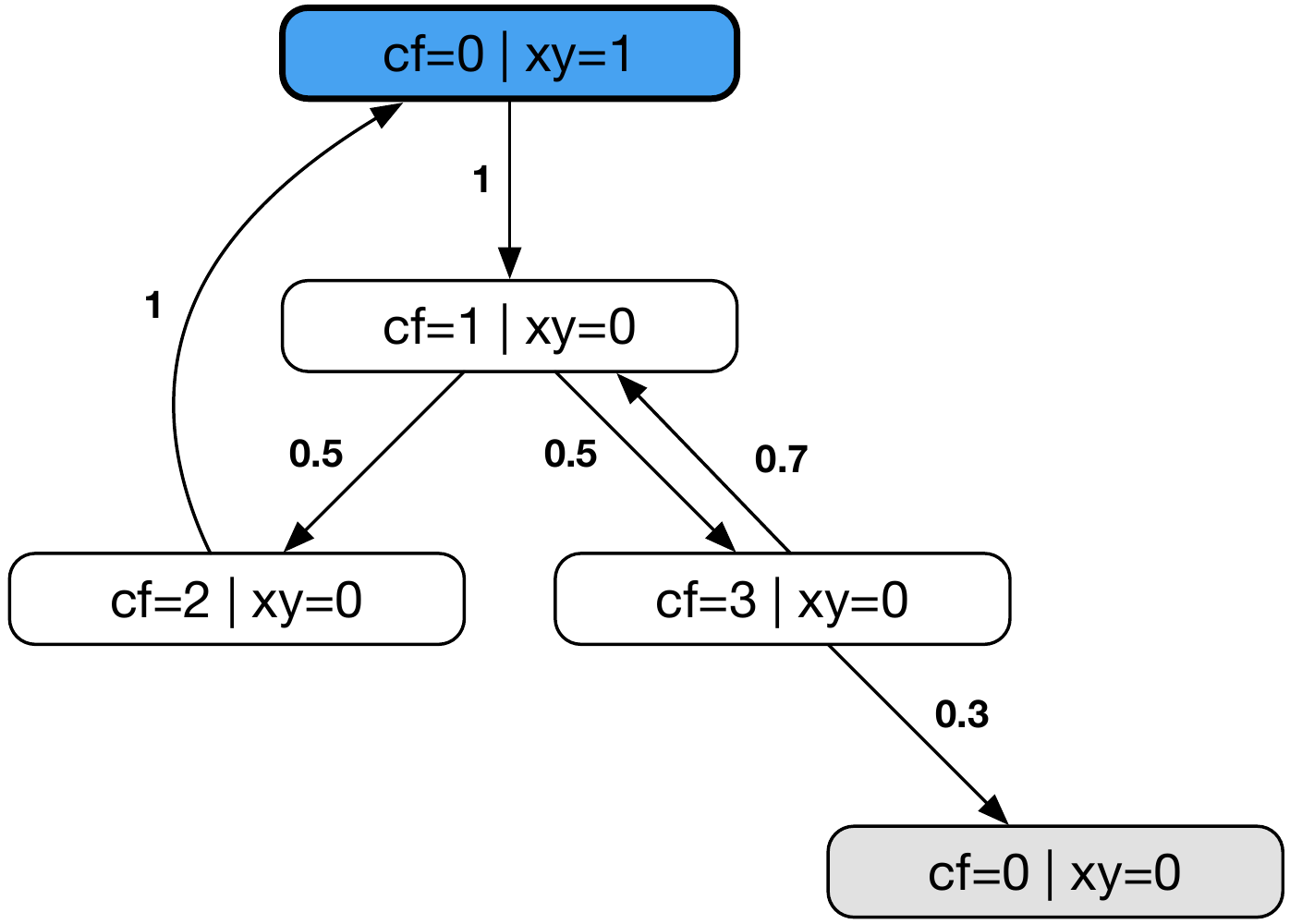}
	\caption{\label{fig:breg}The DTMC semantics of program $\Bsp$ after RAO.}
	\vspace*{-1em}
\end{figure}
The arising DTMC has even one state less than the DTMC obtained after application
of RVO (see Figure~\ref{fig:bres}), showing the potential of even further reductions
on programs through RAO.
\end{example} %

\section{Implementation and Evaluation}\label{sec:evaluation}
We implemented both reduction methods as Python programs that
accept a model in the input language of the prominent 
probabilistic model checker $\prism$~\cite{KNP11}, together with a variable used
as control-flow variable, and a set of variables to be 
excluded from the optimization. 

\begin{table*}[ht]
	\tbl{\label{tab:analysis}Statistics of the reliability analysis experiments}
	{\tabcolsep4pt
	\begin{tabular}{@{}l|rr|rr||rrr|r|r@{}}\toprule 
		reduction&\multicolumn{2}{c|}{model size}&\multicolumn{2}{c||}{reduction}&
				\multicolumn{4}{c|}{times [s]}&speedup\\
		method&states&nodes&states&nodes&IVR&build&analysis&$\sum$\\\colrule
		no& 4.87$\cdot 10^{13}$& 301\,733& -&-&1895.6&708.6&1\,226.2&3\,830.4&-\\
		RVO&1.02$\cdot 10^{11}$& 233\,020& 477.3&1.3&270.4&275.5&915.0&1\,190.4&3.2\\
		RAO&4.30$\cdot 10^{10}$& 121\,426& 1\,133.1&2.5&125.8&82.3&296.8&379.1&10.1\\
	\end{tabular}}
\end{table*}

\paragraph{The VCL model.}
We applied our reduction methods on a simplified version of the aircraft model borrowed 
from the $\simulink$ example set~\cite{aircraft} that itself is based on 
a long-haul passenger aircraft flying at cruising altitude and speed, 
adjusting the fuel flow rate to control the aircraft velocity.
Following the approach of \cite{DDMBJ19}, we generated a model family
that includes options to protect $\simulink$ blocks via redundancy mechanisms
to increase fault tolerance of the velocity control loop (VCL).
Specifically, we considered 13 $\simulink$ blocks protectable with\\
\textbf{comparison:} The block is duplicated and both outputs are compared.
		In case their output differs a dedicated failure state is reached. 
		Otherwise, the output is the one of both blocks.\\
\textbf{voting:} Following the principle of triple modular 
		redundancy principle, the block is triplicated and the 
		output is based on a majority decision.\\
The arising model comprises $3^{13}=1\,594\,323$ different variants
of the VCL. 
Using $\errorpro$~\cite{morozov2019openerrorpro}, we automatically generated
a $\prism$ model on which we applied the reduction methods RVO and RAO presented in this paper.
Here, we modeled errors that could appear during the execution of a $\simulink$ block function, e.g., 
caused by a bit-flip in a CPU. The occurrence of such an error is determined stochastically with 
the probabilities that depend on the commands in the $\simulink$ block functions. 
Even after the reductions, the model could not immediately be constructed by 
$\prism$'s symbolic engine. This was as expected, since also the VCL model considered 
in \cite{DDMBJ19} could not be constructed ad-hoc even given the model presented there
considered only eight block to be protectable.
We hence applied a method called \emph{iterative variable reordering (IVR)}~\cite{DubMorBai20a}
to enable model construction. It is well known that symbolic methods based on
BDDs~\cite{McMillan} as used in $\prism$'s symbolic engine are sensitive to 
the so-called \emph{variable order}. 
The main idea behind IVR
is to iteratively enlarge the system family and perform variable reorderings 
at each step to eventually obtain a suitable variable order for the
whole system family. Specifically, we applied the automated IVR method
presented in~\cite{DubMorBai20a}, following a $\rho$-maximal selection
heuristics with step size four and using the variable reordering implementation
of~\cite{KBCDDKMM18}. Table~\ref{tab:analysis} shows statistics
of the arising model sizes and construction timings. While the number
of nodes used to represent the model symbolically is not significantly reduced,
the explicit state-space representation could be reduced drastically
in 2-3 orders of magnitude.

\paragraph{Reliability analysis.}
We performed a reliability analysis on the constructed VCL model family to
determine the probability of the system to fail within two rounds of execution.
All experiments were carried out on a Linux server system
(2 $\times$ Intel Xeon 
E5-2680 (Octa Core, Sandy Bridge) running at 2.70~GHz with 384~GB of RAM; Turbo Boost 
and Hyper Threading disabled; Debian GNU/Linux 9.1) with a memory bound
of 30~GB of RAM, running the symbolic MTBDD engine of the $\prism$ version 
presented in~\cite{KBCDDKMM18}.
Specifically, we computed $\Pr_{\mathsf{VCL},\pi}(\Diamond^{<2}\fail)$ for every
protection combination $\pi$ using a family-based all-in-one analysis~\cite{DBK15}.
The results range from $0.435$ probability to fail with no protection mechanism and
$0.019$ with all 13 blocks protected. Table~\ref{tab:analysis} shows the timing statistics
of the analysis where an overall speedup of an order of magnitude turns out to be 
possible with our reduction methods.

\paragraph{Comparison to \cite{DDMBJ19}.}
In \cite{DDMBJ19} we proposed to use family-based methods for the analysis of
redundancy systems modeled in $\simulink$, issuing a VCL model family
with eight blocks eligible for protection and four protection mechanisms.
This led to a family size of $4^8=65\,536$. There, we applied preliminary
versions of IVR and RVO in a handcrafted fashion, not exploiting their full potential, e.g.,
leaving out variables that could possibly reset.
In this paper, automated techniques enabled us to perform a reliability
analysis on a much bigger instance with $3^{13}$
family members. %

\section{Discussion and Further Work}\label{sec:conclusion}
To the best of our knowledge, we were the first presenting fully automated
reduction methods for probabilistic programs that purely operate on a syntactic level.
The presented approaches are not only applicable to $\prism$ programs
with a dedicated formalism of control flow, but also to arbitrary $\prism$
programs. However, such optimizations might be not as effective since LRA
heavily relies on the concept of control flows.
In this sense, although we applied our methods on DTMC models only, they
can be also useful for other models expressed in $\prism$'s input language,
e.g., Markov decision processes.
Further, our methods yield bisimilar models for arbitrary
temporal logical formulas expressed over the set of excluded variables. 
Since we
strived towards reliability analysis, we presented the approach in a simpler form
of reward-bounded reachability properties. 
Although RVO uses Chaitin's algorithm, we could use our methods also with 
the linear scan algorithm (LSA)~\cite{poletto1999linear}, e.g.,
to enable fast reduction methods for runtime verification~\cite{FKPQU-RunTimeProb12}.
When bounded model checking~\cite{biere2003bounded} is used as analysis method,
an idea is to adapt LRA such that it considers live ranges up to the desired model-checking bound.
We showed that our methods are effective especially when explicit model representations are used.
As explicit model-checking methods are well known to be faster than symbolic methods when
applied on relatively small models, we expect also such models to benefit from our reduction methods
presented. 
\paragraph{Acknowledgments.}
This work is supported by the DFG through
the Collaborative Research Center TRR 248 
(see {\footnotesize\url{https://perspicuous-computing.science}}, project ID 389792660),
the Cluster of Excellence EXC 2050/1 (CeTI, project ID 390696704, as part of Germany's Excellence Strategy),
the Research Training Groups QuantLA (GRK 1763) and RoSI (GRK 1907), 
projects JA-1559/5-1, BA-1679/11-1, BA-1679/12-1, and the 5G Lab Germany.
\footnotesize
\bibliographystyle{abbrvnat}
\bibliography{main}

\end{document}